\documentclass[final,3p,11pt]{elsarticle}

\usepackage{latexsym}
\usepackage{graphicx}
\usepackage{bbm}
\usepackage{amsmath,amssymb, mathrsfs}
\usepackage[latin1]{inputenc}
\newcommand{\npart}{{N_\textrm{p}}}

\newcommand{\fm}{\textrm{ fm}}
\newcommand{\mb}{\textrm{ mb}}
\newcommand{\ra}{R_{_{\rm A}}}

\newcommand{\bt}{{\boldsymbol{b}_\perp}}

\newcommand{\nr}[1]{(\ref{#1})} 
\newcommand{\ud}{\mathrm{d}}
\newcommand{\fig}{fig.~}
\newcommand{\figs}{figs.~}
\newcommand{\eq}{eq.~}
\newcommand{\se}{sec.~}
\newcommand{\eqs}{eqs.~}

\newcommand{\nf}{{N_\mathrm{F}}}
\newcommand{\nb}{{N_\mathrm{B}}}

\newcommand{\rmF}{{\mathrm{F}}}
\newcommand{\rmB}{{\mathrm{B}}}
\newcommand{\rmR}{{\mathrm{R}}}
\newcommand{\nref}{{N_\mathrm{R}}}
\newcommand{\nfbar}{{\overline{N}_\mathrm{F}}}
\newcommand{\nbbar}{{\overline{N}_\mathrm{B}}}
\newcommand{\nbar}{{\overline{N}}}
\newcommand{\nrefbar}{{\overline{N}_\mathrm{R}}}

\newcommand{\sff}{{\sigma^2_{_\mathrm{FF}}}}
\newcommand{\sbb}{{\sigma^2_{_\mathrm{BB}}}}
\newcommand{\srr}{{\sigma^2_{_\mathrm{RR}}}}
\newcommand{\sfr}{{\sigma^2_{_\mathrm{FR}}}}
\newcommand{\sbr}{{\sigma^2_{_\mathrm{BR}}}}
\newcommand{\sfb}{{\sigma^2_{_\mathrm{FB}}}}

\newcommand{\rfr}{{R_{_\mathrm{FR}}}}
\newcommand{\rbr}{{R_{_\mathrm{BR}}}}
\newcommand{\rfb}{{R_{_\mathrm{BF}}}}
\newcommand{\rbf}{{R_{_\mathrm{BF}}}}

\begin{document}

\begin{frontmatter}

\title{Long range rapidity correlations as seen in the STAR experiment}
\author[jyu,cea]{T. Lappi}
\author[bnl]{L. McLerran}
\address[jyu]{Department of Physics, %
 P.O. Box 35, 40014 University of Jyv\"askyl\"a, Finland}
\address[cea]{Institut de Physique Th\'eorique, %
B\^at. 774, CEA/DSM/Saclay, 91191 Gif-sur-Yvette, France}
\address[bnl]{Physics Department and Riken-BNL Center, Brookhaven
National Laboratory, Upton, NY 11973, USA}

\begin{abstract}
We analyze  long range rapidity correlations observed in the STAR experiment at RHIC.  Our goal is
to extract properties of the two particle correlation matrix, accounting for the 
analysis method of the STAR experiment.  We find a surprisingly large correlation 
strength for central  collisions of gold nuclei at highest RHIC energies.  We argue that
 such correlations  cannot be the result of impact parameter fluctuations.
\end{abstract}

\begin{keyword}
%% keywords here, in the form: keyword \sep keyword

%% PACS codes here, in the form: \PACS code \sep code

%% MSC codes here, in the form: \MSC code \sep code
%% or \MSC[2008] code \sep code (2000 is the default)

\end{keyword}

\end{frontmatter}

\section{Introduction}
The STAR experiment at RHIC has reported preliminary observations of forward backward 
correlations as a function of the centrality of the 
collision~\cite{Abelev:2009dq,Tarnowsky:2008am}.
The reported forward-backward correlation shows a rapid increase in strength as a
 function of centrality, and appears to have a strength which cannot be explained
 by a superposition of $pp$ interactions.  It has been suggested that such 
correlations are a consequence of the Color Glass Condensate induced Glasma 
produced in the early stages of heavy ion 
collisions~\cite{McLerran:1993ni,McLerran:1993ka,Kovchegov:1996ty,Kovner:1995ja,%
Kovner:1995ts,JalilianMarian:1996xn,JalilianMarian:1997jx,JalilianMarian:1997gr,%
Iancu:2000hn,Iancu:2001ad,Ferreiro:2001qy,Krasnitz:1999wc,Krasnitz:1998ns,%
Lappi:2003bi,Lappi:2006fp,Kovchegov:1999ep,Armesto:2006bv,Fukushima:2008ya}  An alternative 
but quite similar explanation is provided by parton 
percolation~\cite{Armesto:1996kt,Braun:1997ch,Cunqueiro:2006xe}.
The long range correlations may be the origin of the ridge phenomenon measured at RHIC,
\cite{Adams:2004pa,Adams:2005ka,Adams:2005aw,Putschke:2007mi,Daugherity:2008su,Adare:2008cq,%
McCumber:2008id,Wenger:2008ts} that may also be explained as arising from color 
electric and magnetic flux tubes originating in the
Glasma~\cite{Dumitru:2008wn,Gavin:2008ev,Dusling:2009ar,Gelis:2009wh}.

The results for the forward-backward correlations strength measured in the 
STAR experiment 
are reported as a function of centrality bin.  One might be worried that if one 
averages over all the events in a centrality bin, correlations
would be generated by the  different impact parameters (or numbers of wounded nucleons)
possible within such a  bin; this is essentially the mechanism proposed 
in~\cite{Konchakovski:2008cf,Bzdak:2009xq,Bzdak:2009bc,Bzdak:2009st}.  We expect 
the charged multiplicity to be strongly correlated with impact parameter and if the 
impact parameter itself can have significant variation within a fixed centrality bin,
then spurious correlations whose only origin is the geometry of the 
collision would be generated.

The STAR analysis is however more sophisticated.  
For each event one measures also a reference multiplicity
in a relatively wide central rapidity interval.   
Correlations and fluctuations of the forward and backward multiplicities
are then measured separately for each reference  multiplicity.
The correlations 
and fluctuations reported are average values of these measurements over each 
centrality  bin.  One might expect that such a procedure is relatively 
insensitive to impact  parameter correlations.  

One purpose of this paper is to show how to extract values of the multiplicity 
correlation function including the constraints actually applied by the the STAR 
experiment.  After setting up some notations in \se\ref{sec:rapcorr}
we argue in \se\ref{sec:gauss} that, 
within a Gaussian approximation for the fluctuations, there is a bound on 
the ratio of forward backward correlated fluctuations to forward fluctuations,
\begin{equation}
 b = \frac{ \langle \nf \nb \rangle_{\nref} - 
 \langle \nf \rangle_{\nref} \langle \nb \rangle_{\nref}} 
{ \langle \nf^2\rangle_{\nref} 
-  \left\langle \nf \right\rangle^2_{\nref}} .
\end{equation}
Here, $\nf$ is the multiplicity measured in some forward bin of rapidity, and 
$\nb$ the value in a backwards bin, and the subscript $\nref$ indicates that 
the averages are taken for fixed values of the reference multiplicity.
 In the STAR experiment the pseudorapidity windows are at 
symmetric values of pseudorapidity around $\eta =0$.  When the 
forward and backward rapidity windows are smaller than the reference
window and farther from each other than from the reference window,  we argue on 
quite general grounds that $b < 1/2$.  This is perhaps violated by non-linear
 effects that are of order $1/\langle \nf \rangle$, effects that could be several percent 
for the most central collisions.  If $b = 1/2$, the multiplicities are maximally 
correlated.  This bound is derived making, besides this Gaussian approximation,
only minimal assumptions about the correlations and does not depend on
the detailed mechanism for generating them.
The STAR data is consistent with this bound except for the most 
central events of Au-Au collisions at the highest beam energy, where there are 
small violations.  
We then analyse the effect of the impact parameter induced correlations 
of the type suggested  
in Ref.~\cite{Konchakovski:2008cf,Bzdak:2009xq,Bzdak:2009bc,Bzdak:2009st} 
in \se\ref{sec:torrieri} and find that they
are not sufficient to 
explain the large correlation seen in the STAR data.  In \se\ref{sec:corr}
we then propose a simple parametrization to add intrinsic long range correlations
to the impact parameter fluctuations.

Our results are: The correlations measured in STAR do not appear to be entirely 
associated with impact parameter fluctuations.  
The measured value of $b$ is near its maximally allowed value for central Au-Au 
collisions at RHIC energy.  Indeed for the highest energy 
and most central Au-Au collisions it is larger than the bound $b < 1/2$.
We find that a crucial requirement to obtain a $b$ close to this limit are
large fluctuations in the reference multiplicity.

\section{The rapidity correlation function}
\label{sec:rapcorr}

The quantity that we wish to extract information from experiment is the rapidity 
correlation function.
We define it as
\begin{equation}\label{eq:defC}
C(\eta,\eta') \equiv 
\left\langle  \frac{\ud N}{\ud \eta} \frac{\ud N}{\ud \eta'} \right\rangle
- \left\langle  \frac{\ud N}{\ud \eta}\right\rangle 
   \left\langle \frac{\ud N}{\ud \eta'} \right\rangle
\end{equation}
This can be  decomposed  into a local piece that corresponds to Poissonian 
fluctuations in the particle number and a long range correlaion
\begin{equation}\label{eq:generalC}
C(\eta-\eta') =  \delta(\eta-\eta') \left\langle  \frac{\ud N}{\ud \eta} \right\rangle 
 + K(\eta-\eta') 
 \left\langle \frac{\ud N}{\ud \eta} \right\rangle 
 \left\langle \frac{\ud N}{\ud \eta'} \right\rangle .
\end{equation}
The $\delta(\eta-\eta')$ piece is often (e.g. \cite{Giovannini:1985mz,Adler:2007fj})
absorbed into the definition of the correlation
function itself (the l.h.s. of \eq\nr{eq:defC}), but it will be convenient for the 
following discussion to keep it on the r.h.s. as  part of $C(\eta-\eta')$. 

Some information about the function $K$ can be found by measuring local 
multiplicity fluctuations.   Integrating
\eq\nr{eq:generalC} twice over a pseudorapidity interval $\Delta \eta$  
centered at some rapidity $\eta$ and assuming   that $\Delta \eta$ is sufficiently small
that the pseudorapidity dependence of the multiplicity in this interval  can
be neglected,  we obtain 
\begin{equation}\label{eq:negbin}
\langle n^2\rangle - \langle n\rangle^2
= \langle n \rangle + \frac{\langle n\rangle^2}{k}.
\end{equation}
This result is identical to the result for obtained a negative binomial distribution, and 
we can  identify the $k$-parameter as
\begin{equation}\label{eq:negbink}
\frac{1}{k} = \int_{\eta -\Delta \eta/2}^{\eta +\Delta \eta/2} \ud\eta ~\ud\eta' ~  K(\eta-\eta').
\end{equation}
(The negative binomial distribution has further consequences for higher moments of the multiplicity
distribution as well.) 
This relation can be used to gain information on the rapidity correlation for small rapidity differences from 
measurements of the multiplicity distribution in different 
$\Delta \eta$-intervals~\cite{Giovannini:1985mz,Adler:2007fj}

Of central interest in this paper is the behavior of the correlation function $C$ at 
large rapidity differences.  In particular, we shall concentrate on the case
 studied by the STAR 
collaboration~\cite{Abelev:2009dq,Tarnowsky:2008am}. This analysis 
measures the correlation between charged particle multiplicities
in forward (``F'') and backward (``B'') rapidities.
Specifically, 
pseudo-rapidity windows of width $\delta=0.2$, are situated symmetrically 
around midrapidity within the STAR TPC acceptance $|\eta|<1$. The data is
divided into 10\% centrality bins and the correlation coefficient 
between the forward and backward multiplicities is measured within the
centrality bin.  In order to 
eliminate the effect of centrality fluctuations within one bin, there is an 
additional crucial twist in the analysis. One also measures the 
multiplicity in a \emph{third}
reference (``R'') rapidity interval of width $\delta_\mathrm{R}=1.0$
that does not overlap with the ``F'' and ``B'' windows. The 
variances and the covariance of the ``F'' and ``B'' windows are then 
determined separately for each reference multiplicity $N_\mathrm{R}$, 
the  motivation being that a fixed $N_\mathrm{R}$ selects, with a good 
accuracy, events with a fixed impact parameter. Although the data is
presented for rapidity intervals in different locations, it is good 
to keep in mind a typical configuration, with rapidity windows
$-1<\eta<-0.8$ (``B''), $-0.5<\eta<0.5$ (``R'') and $0.8<\eta<1$ (``F'').

\section{Gaussian approximation}
\label{sec:gauss}

As we have discussed, in order to analyze the STAR results, we shall 
have to study the case
of three correlated multiplicities, $\nf$, $\nb$ and $\nref$.
We shall first compute such fluctuations in a Gaussian approximation for 
the probability distribution of these three variables within all the events in 
a centrality bin.  This should be a good approximation
when considering fluctuations around an average multiplicity, so long as that 
average multiplicity is large.  This is a consequence of the central limit theorem, 
and in \se\ref{sec:torrieri} we shall see explicitly how it follows directly
 from a stationary phase 
approximation to expressions for the multiplicity
fluctuations in a simple model.  The corrections to this approximations 
should be of order $1/N$, where $N$ is  a multiplicity in a bin.  For the
STAR experiment, and central events, this should
be valid to a few percent, since the typical value of $N_{F,B} \sim 100$ 
for the forward  or backward multiplicity.  However, in events with 
centrality around  50\%, $N_{F,B} \sim 20$, 
so one can expect 10-20\% corrections.

We assume that there is some requirement on the centrality of the 
collision that fixes the average
values $\langle \nf \rangle$, $\langle \nb \rangle$, and $\langle\nref \rangle$. 
We define $\Delta_U = N_U - \langle N_U \rangle$, where $U$
is $\rmF$, $\rmB$ or $\rmR$, so that the 
normalized probability distribution is 
\begin{equation}\label{eq:gauss}
   P(\nf,\nb;\nref) = {1 \over {(2\pi)^{3/2} \det \Sigma }} 
\exp \bigg[ -{1 \over 2} \Delta_U \Sigma^{-1}_{UV} \Delta_V \bigg]
\end{equation}
In this equation
\begin{eqnarray}
\label{eq:defsigma}
\Sigma_{UV}   \equiv 
   \sigma^2_{UV}   &  = &   
\langle \Delta_U  \Delta_V \rangle
 =  \int_{U}\ud \eta \int_{V} \ud \eta' ~C(\eta-\eta') \quad ,
 \quad \quad U,V =  \rmF , \rmB , \rmR.
\end{eqnarray}
In all that follows, we will assume that the forward and backward windows 
are chosen to be centered at rapidity values symmetric around $\eta = 0$, 
and that the reference multiplicity is chosen by summing
over rapidity values that are also symmetrically displaced. Since 
$\Sigma_{UV} = \Sigma_{VU}$ by construction this means that
\begin{align}
          \Sigma_{\rmF  \rmB} & =  \Sigma_{\rmB \rmF}  \\
  \Sigma_{\rmF \rmR} = \Sigma_{\rmB \rmR} & =  \Sigma_{\rmB \rmR} = \Sigma_{\rmR \rmB}
\end{align}             
Other than the Gaussian
form and the symmetry above, we shall in this section make 
no assumptions about the physical origin
or strength of the correlation; except that it is a decreasing function of 
the rapidity separation between the measured multiplicities.
We also introduce the \emph{correlation coefficient} between two
multiplicities with the conventional definition
\begin{equation}
R_{UV} = \frac{\Sigma_{UV}}{\sqrt{\Sigma_{UU} \Sigma_{VV}}}.
\end{equation}
The correlation coefficient is mathematically
restricted (due to the Cauchy-Schwarz inequality) to values between $-1$ and $1$.
Without fixing the reference multiplicity, the covariance $D_{bf}$, variance
$D_{ff}$ and correlation coefficient $b$  used 
in~\cite{Abelev:2009dq,Tarnowsky:2008am} would be the same
as $\sfb$ $\sff$ and
$\rfb$. We shall however reserve the notation $D_{bf}$
$D_{ff}$ and $b$ to the quantities at fixed $\nref$ that are actually measured.
The probability distribution \nr{eq:gauss} involves $\Sigma^{-1}_{UV}$, 
the inverse of the 
correlation matrix $\Sigma$. As a $3\times3$ matrix $\Sigma$ is easily inverted to give
\begin{equation}\label{eq:sigmainv}
\Sigma^{-1} = 
\frac{1}{\det \Sigma}
\left( 
\begin{array}{ccc}
\sbb\srr - \left(\sbr\right)^2 & \sbr\sfr - \sfb \srr & \sbr \sfb - \sbb \sfr \\[1ex]
\sbr \sfr - \sfb \srr &  \sff \srr - \left(\sfr\right)^2  & \sfb \sfr  -\sbr \sff  \\[1ex]
\sbr \sfb - \sbb \sfr & \sfb \sfr -\sbr \sff & \sbb \sff- \left(\sfb\right)^2  \\
\end{array}
\right).
\end{equation}

When the reference multiplicity is not measured, the probability
distribution  can be reduced to the double distribution for
$\Delta_\rmF,\Delta_\rmB$ by integrating over $\Delta_\rmR$
\begin{eqnarray}
P(\Delta_\rmF,\Delta_\rmB) 
&=& 
 \int \ud \Delta_\rmR P(\Delta_\rmF,\Delta_\rmB,\Delta_\rmR)  \nonumber \\
& =& 
\frac{1}{(2\pi)^{3/2}\det {\Sigma_2}}
\exp \left[ -\frac{ \Delta_U {\Sigma_2^{-1}}_{UV} \Delta_V}{2} \right] \quad, 
\quad \quad U,V = F,B
\end{eqnarray}
where ${\Sigma_2}^{-1}$ is the inverse of the $2\times2$ correlation matrix
\begin{equation}
{\Sigma_2}  = \left( 
\begin{array}{cc}
\sff & \sfb \\[1ex]
\sfb & \sbb  \\
\end{array}
\right).
\end{equation}
In our case, however, we want the  probability 
distribution of $\Delta_\rmF,\Delta_\rmB$ for a fixed $\Delta_\rmR$,
not integrated over all values of $\Delta_\rmR$. In this 
case the (conditional on $\Delta_\rmR$)
 probability distribution for $\Delta_\rmF,\Delta_\rmB$
is given by
\begin{eqnarray}
P(\Delta_\rmF,\Delta_\rmB \, | \, \Delta_\rmR) &  =  & 
\frac{P(\Delta_\rmF,\Delta_\rmB,\Delta_\rmR) }{P(\Delta_\rmR)}  \\
 & = & 
\frac{1}{2\pi \det \widetilde{\Sigma}}
\exp \bigg[ - \frac{ \Delta_U \widetilde{\Sigma}^{-1}_{UV} \Delta_V}{2} 
- {\Sigma^{-1}}_{U\rmR} \Delta_U \Delta_\rmR 
- \frac{ {\Sigma^{-1}}_{\rmR \rmR} - (\Sigma_{\rmR\rmR})^{-1}}{2} \Delta_\rmR^2
\bigg]  \nonumber \\
& & \quad \quad U,V = \rmF , \rmB. 
\nonumber
\end{eqnarray}
The presence of mixed terms in 
$\Delta_\rmF \Delta_\rmR, \Delta_\rmB \Delta_\rmR $
means that the expectation values of $\Delta_\rmF, \Delta_\rmB$
are shifted from zero and proportional to $\Delta_\rmR$. This 
is natural, since if one has an event with exceptionally large 
$\nref$ one also expects a large $\nf$ and $\nb$ because
the multiplicities are correlated.
The term in $\Delta_\rmR^2$ assures the correct normalization 
$\int \ud \Delta_\rmF \ud \Delta_\rmB P_\rmR(\Delta_\rmF,\Delta_\rmB)=1$.
The fluctuations and correlations of $\Delta_\rmF, \Delta_\rmB$ are determined by 
the coefficients of the quadratic part. They are described by
the reduced correlation matrix $\widetilde{\Sigma}$, obtained by taking
the $\rmF,\rmB$-elements of $\Sigma^{-1}$ from \eq\nr{eq:sigmainv}
\begin{equation}\label{eq:sigmatildeinv}
\widetilde{\Sigma}^{-1} = 
\frac{1}{\det \Sigma}
\left(
\begin{array}{cc}
\sbb\srr - \left(\sbr\right)^2 & \sbr\sfr - \sfb \srr  \\[1ex]
\sbr \sfr - \sfb \srr &  \sff \srr - \left(\sfr\right)^2    \\
\end{array}
\right).
\end{equation}
To get the correlation matrix of $\Delta_\rmF,\Delta_\rmB$ for a fixed 
$\Delta_\rmR$ we must then again invert this matrix to get
\begin{equation}\label{eq:sigmatilde}
\widetilde{\Sigma} = \left( 
\begin{array}{cc}
\sff - \frac{\left(\sfr\right)^2}{\srr} & \sfb - \frac{\sbr \sfr}{\srr} \\[2ex]
\sfb - \frac{\sbr \sfr}{\srr} & \sbb - \frac{\left(\sbr\right)^2}{\srr}  \\
\end{array}
\right).
\end{equation}

From this we can read off correlation coefficients that correspond to the 
measured observables
\begin{eqnarray}
D_{ff} &=& \widetilde{\Sigma}_{\rmF \rmF} = \sff - \frac{\left(\sfr\right)^2}{\srr}
\\
D_{bf} &=& \widetilde{\Sigma}_{\rmB \rmF} = \sfb - \frac{\sbr \sfr}{\srr}
\\
b      &=& \frac{D_{bf}}{\sqrt{D_{ff} D_{bb}}} = 
\frac{\rbf -  \rbr \rfr}{ \sqrt{\left(1 - \left(\rbr\right)^2\right)
\left( 1- \left(\rfr\right)^2\right)} }.
\label{eq:b}
\end{eqnarray}

The shifts in the expectation values of the $\Delta$'s for a fixed 
reference multiplicity can be computed in terms of the 
matrix elements of $\Sigma$ as
\begin{equation}
\langle \Delta_{\rmF,\rmB} \rangle_{\nref} = \frac{\sfr}{\srr} \Delta_\rmR 
\end{equation}
The information in the correlation between the reference multiplicity and the forward and backward
multiplicity, as well as the fluctuations in the reference multiplicity itself contain some
non-trivial information about the correlations.

Equation \nr{eq:b} is the central result of this section. It describes
how measurement of the forward-backward correlation is modified by the 
fact that the multiplicities are observed for a fixed reference 
multiplicity, which in turn is correlated with the forward and backward 
multiplicities. We shall now turn to analysing its consequences using
some very general assumptions on the two particle correlation 
\nr{eq:defC}.
The corrections caused by the fixed reference multiplicity have a
very intuitive meaning. For a fixed reference multiplicity the 
$\Delta_{\rmF,\rmB}$ fluctuate less,
 $\widetilde{\Sigma}_{\rmF \rmF} < \sff $. This is due to the correlation 
between the reference and $\rmF,\rmB$ multiplicities and the effect goes away
when $\rfr \to 0$. The modification also decreases when the reference multiplicity
has larger (uncorrelated) fluctuations $\srr$.

In the limit of very strong correlations, the correlation coefficients
$R$ are all very large. This is the case in heavy ion collisions when the
centrality bins are taken to be very wide so that fluctuations and correlations
are dominated by impact parameter fluctuations within the centrality bin.
Let us first assume, for simplicity, that all the correlation 
coefficients are equal, $\rfb = \rfr = \rbr = R$. 
Now the conditional correlation coefficient \nr{eq:b} is given by
\begin{equation}
b = \frac{R-R^2}{1-R^2} = \frac{R}{1+R}.
\end{equation}
Remembering that $R \leq 1$ this lead to the upper limit $b<1/2$.

Consider now the typical experimental situation mentioned above, 
with rapidity windows
$-1<\eta<-0.8$ (``B''), $-0.5<\eta<0.5$ (``R'') and $0.8<\eta<1$ (``F'').
We shall only assume that the correlation function $C(\eta,\eta')$
is boost-invariant  and symmetric, i.e. a function of $|\eta-\eta'|$
only. Any reasonable correlation function $C(|\eta-\eta'|)$ will 
decrease as a function of $|\eta-\eta'|$. 
These very general assumptions inserted into \eq\nr{eq:defsigma} lead 
to simple inequalities for the correlation matrix:
\begin{equation}\label{eq:estvar}
\frac{1}{\delta_\rmR^2} \srr \leq \frac{1}{\delta^2} \sff = \frac{1}{\delta^2} \sbb
\textrm{ for } \delta_\rmR > \delta.
\end{equation}
Because in our typical configuration the $\rmF,\rmB$-windows are closer to the
reference rapidity window than to each other, the average value of the correlation
function between the $\rmF,\rmB$ windows and the reference window is also larger. 
This leads to the estimate
\begin{equation}\label{eq:estcov}
\frac{1}{\delta_\rmR \delta} \sfr \geq \frac{1}{\delta^2} \sfb.
\end{equation}
Together \eqs\nr{eq:estvar} and~\nr{eq:estcov} imply that $\rfr=\rbr \geq \rfb$. 
Because the constrained correlation coefficient $b$ 
in \eq\nr{eq:b} is a monotonously \emph{decreasing} function of $\rfr=\rbr$, we 
then get the upper limit
\begin{equation}
b = \frac{\rbf -\rfr^2}{1-\rfr^2} \leq  \frac{\rbf -\rbf^2}{1-\rbf^2} 
=  \frac{\rbf}{1+\rbf} \leq 0.5  \textrm{ for } \delta_\rmR > \delta.
\end{equation}
The preliminary STAR results~\cite{Abelev:2009dq,Tarnowsky:2008am,Srivastava:2007ei}
violate this bound derived from very general assumptions, and at the moment
we  see no compelling explanation for this discrepancy.
Note also that for the configurations where the $\rmF,\rmB$-windows are close to each 
other, their distance from each other can be less than that to the reference 
rapidity region. When $\rmF,\rmB$ windows are closer than
0.6 units in rapidity, the reference window is determined as $0.5< |\eta|<1.0$ 
and \eq\nr{eq:estcov} no longer strictly applies. In this region the
observed value of $b$ does indeed tend to be slightly larger than for large rapidity separations.

\section{Poissonian and impact parameter fluctuation model}
\label{sec:torrieri}

\subsection{The model}

For an illustration let us consider the toy model introduced in
Ref.~\cite{Konchakovski:2008cf}, where the authors claim
 that the impact parameter fluctuations within the centrality bin would explain
most of the STAR correlation measurement. The model includes a short
range Poisson correlation and the fluctuation of the collision geometry 
via a Monte Carlo Glauber (MCG) model. In practice the charged particles
produced in a nucleus-nucleus collision are assumed to be a superposition
of particle production from each participant nucleon:
\begin{equation}\label{eq:torrierimodel}
N_\textrm{ch} = \sum_{i=1}^{\npart} n_i, 
\end{equation}
where $\npart$ is the number of participant nucleons. The variables
$n_i$ are independent and distributed according to a Poisson distribution 
with mean $\bar{n}$. These particles are then distributed in pseudorapidity 
with a gaussian distribution so that
\begin{equation}\label{eq:gausseta}
 \left\langle  \frac{\ud N}{\ud \eta}\right\rangle  = 
\left\langle \npart  \right\rangle \frac{\bar{n}}{\sqrt{2 \pi \sigma_\eta}}
e^{-\frac{\eta^2}{2 \sigma_\eta^2}}.
\end{equation}

\subsection{Gaussian approximation}

Knowing that the only correlations in this toy model are the local one leading to 
the Poisson distribution and an infinite range one from the fluctuating number of 
participant nucleons, one can immediately write down\footnote{
We are neglecting quantities that are higher order in the multiplicity fluctuations.
For example one is approximating $\langle n_i^2 \rangle \approx \langle n_i \rangle^2$
in the second term, since it is already proportional to the relative variance
$\sigma^2_\textrm{p}/\left\langle \npart  \right\rangle^2$
which can be assumed to be small.} the correlation function of the
model:
\begin{equation}
C(\eta,\eta') = \delta(\eta-\eta') \left\langle  \frac{\ud N}{\ud \eta} \right\rangle 
 +  \frac{\sigma^2_\textrm{p}}{\left\langle \npart  \right\rangle^2 }
 \left\langle \frac{\ud N}{\ud \eta} \right\rangle 
 \left\langle \frac{\ud N}{\ud \eta'} \right\rangle ,
\end{equation}
where $\sigma^2_\textrm{p}$ is the variance of the number of participants in
the given centrality bin. It is easy to see that this correlation function,
when integrated over some intervals in the pseudorapidities $\eta$ and
$\eta'$, reproduces the variance of $N_\textrm{ch}$ defined by 
\eq\nr{eq:torrierimodel}.

 Integrating this over the rapidity windows gives then
\begin{eqnarray}\label{eq:torrierivar}
\sff &=& \langle N_\rmF \rangle +   \frac{\sigma^2_\textrm{p}}
	{\left\langle \npart  \right\rangle^2 } \langle N_\rmF \rangle^2
\\ \label{eq:torriericov}
\sfb &=& \frac{\sigma^2_\textrm{p}} {\left\langle \npart  \right\rangle^2 } \langle N_\rmF \rangle \langle N_\rmB \rangle
\end{eqnarray}
and similarly for the other interval combinations. This gives correlation coefficients
\begin{equation}
\rfb = \frac{1}{
\sqrt{1 +
\frac{\left\langle \npart  \right\rangle^2 }{\sigma^2_\textrm{p}}
\frac{1}{\langle N_\rmF \rangle} }
\sqrt{1 +
\frac{\left\langle \npart  \right\rangle^2 }{\sigma^2_\textrm{p}} 
\frac{1}{\langle N_\rmB \rangle} }
}
\end{equation}
and correspondigly for $\rfr$ and $\rfb$.

To get a rough estimate of the numbers we shall neglect the $\eta$-dependence in 
\eq\nr{eq:gausseta} and approximate $\tilde{n}\equiv \bar{n}/\sqrt{2 \pi \sigma_\eta} \approx 2$
(this is the charged multiplicity per unit rapidity and per participant).
 Taking the rapidity intervals
$\delta$ and $\delta_\rmR$ for the different windows  we then get
\begin{eqnarray} \label{eq:torrierifb}
\rfb &=& \frac{1}{1+ \frac{1}{\omega_\textrm{p} \bar{n} \delta}}
\\ \label{eq:torrierifr}
\rfr &=& \frac{1}{
\sqrt{1+ \frac{1}{\omega_\textrm{p} \tilde{n} \delta}}
\sqrt{1+ \frac{1}{\omega_\textrm{p} \tilde{n} \delta_\rmR}}
},
\end{eqnarray}
where we have, following~\cite{Konchakovski:2008cf}, denoted
$\omega_\textrm{p} \equiv \sigma^2_\textrm{p}/\left\langle \npart \right\rangle$.
We then take from \cite{Konchakovski:2008cf} the values $\omega_\textrm{p} \approx 2.5$
for the 0--10\% bin and $\omega_\textrm{p} \approx 1.3$ for a midcentral 40--50\%.
bin.
This gives  $\rfb \approx 0.5$ and $\rfr \approx 0.65$ for the more central bin
and $\rfb \approx 0.34$ and $\rfr \approx 0.5$. These values of $\rfb$ are the 
correlation coefficients quoted in \cite{Konchakovski:2008cf}. As we have discussed,
however, they do not correspond to the actual measured quantity. 

Only at this point need we evoke the Gaussian approximation for the probability
distribution. If we assume that the probability distribution is Gaussian,
we can use  \eq\nr{eq:b} to evaluate the $b$ parameter from the 
correlation coefficients \nr{eq:torrierifb} and \nr{eq:torrierifr} of the model.
Putting the numbers estimated above into \eq\nr{eq:b} gives
$b\approx 0.14$ for the central and $b\approx 0.12$ for the midcentral bin. These
values are clearly far from the experimental result.

\subsection{Beyond the Gaussian approximation}

Although it seems unlikely, based on the result we just obtained, that this simple toy
model would come near to explaining the experimental data, we shall still 
continue analysing it further in order to understand the effect of  the Gaussian approximation. 
For this it is instructive to start over from the definition of the model as a probability 
distribution.
We shall again neglect the $\eta$-dependence of the single particle spectrum
to avoid encumbering our notations. The model is defined by a fluctuating
number of participant nucleons $\npart$, which gives the parameter for the 
Poisson-distributions of the measured multiplicities $N_\rmF, \ N_\rmB, \ N_\rmR$.
We can therefore write down the probability distribution that characterizes the model
as follows
\begin{equation}\label{eq:deftorrieri}
P(\npart,\nf, \nb, \nref) = 
P_\textrm{MCG}(\npart) 
\left( \frac{ \left( \delta \tilde{n} \npart \right)^{N_\rmF} e^{-\delta \tilde{n} \npart}}{N_\rmF !}\right)
\left( \frac{ \left( \delta \tilde{n} \npart \right)^{N_\rmB} e^{-\delta \tilde{n} \npart}}{N_\rmB !}\right)
\left( \frac{ \left( \delta_\rmR  \tilde{n} \npart \right)^{N_\rmR} e^{-\delta_\rmR  \tilde{n} \npart}}{N_\rmR !}\right).
\end{equation}
Here $P_\textrm{MCG}(\npart)$ is the probability distribution of events with different
numbers of participant nucleons, presumably to be obtained from a Monte Carlo Glauber calculation.
The first step in our approximation is to replace the Poissonian distributions by Gaussians
in $\nf, \ \nb, \ \nref$, which should be pretty safe as long as these multiplicities 
are all large enough\footnote{We are simultaneously 
approximating the discrete variables $\nf, \ \nb, \ \nref$ by continuous ones}. We then obtain
\begin{equation}\label{eq:torrierigauss}
P(\npart,\nf, \nb, \nref) = 
\frac{P_\textrm{MCG}(\npart)}{(2 \pi \tilde{n} \npart)^{3/2}\delta \sqrt{\delta_\rmR}} \exp \left[
-\frac{ \left( \nf - \delta \tilde{n} \npart\right)^2 }{2 \delta \tilde{n} \npart}
-\frac{ \left( \nb - \delta \tilde{n} \npart\right)^2 }{2 \delta \tilde{n} \npart}
-\frac{ \left( \nref - \delta_\rmR \tilde{n} \npart\right)^2 }{2 \delta_\rmR \tilde{n} \npart}
\right].
\end{equation}
The next step in our approximation introduces a larger error. In the experimental analysis
the centrality class is defined by events where the reference multiplicity $\nref$ falls 
between a given lower and upper bound. Since the number of participants is not a directly 
experimentally observable quantity, it cannot be used as a selection criterion. We shall,
however, assume that the centrality bin is defined by a Gaussian
distribution of $\npart$ with some variance $\sigma^2_\textrm{p}$:
\begin{equation}
P_\textrm{MCG}(\npart) \approx \frac{1}{\sqrt{2 \pi \sigma^2_\textrm{p}}}
\exp \left[-\frac{(\npart -\langle \npart \rangle)^2}{2  \sigma^2_\textrm{p}} \right].
\end{equation}
The remaining approximation is to replace $\npart$  by $\langle \npart \rangle$
 in the variances 
(but not the means!) of the Gaussians in  
$\nf, \ \nb, \ \nref$\footnote{This is equivalent to
the approximation $\langle n_i^2 \rangle \approx \langle n_i \rangle^2$ in the variance of 
$\ud N/\ud \eta$ mentioned in an earlier footnote.}. This has the result of turning our probability 
distribution into a Gaussian in all four of its variables:
\begin{equation}\label{eq:gausstorrieri}
P(\npart,\nf, \nb, \nref) = 
\frac{
\exp \left[
-\frac{(\npart -\langle \npart \rangle)^2}{2  \sigma^2_\textrm{p}}
-\frac{ \left( \nf - \delta \tilde{n} \npart\right)^2 }{2 \delta \tilde{n} \langle \npart \rangle}
-\frac{ \left( \nb - \delta \tilde{n} \npart\right)^2 }{2 \delta \tilde{n} \langle \npart \rangle}
-\frac{ \left( \nref - \delta_\rmR \tilde{n} \npart\right)^2 }{2 \delta_\rmR \tilde{n} 
\langle \npart \rangle}
\right]
}
{(2 \pi)^2 (\tilde{n} \langle \npart \rangle)^{3/2}\delta \sqrt{\delta_\rmR} \sigma_\textrm{p}}.
\end{equation}

\begin{figure}
\begin{center}\includegraphics[width=0.6\textwidth]{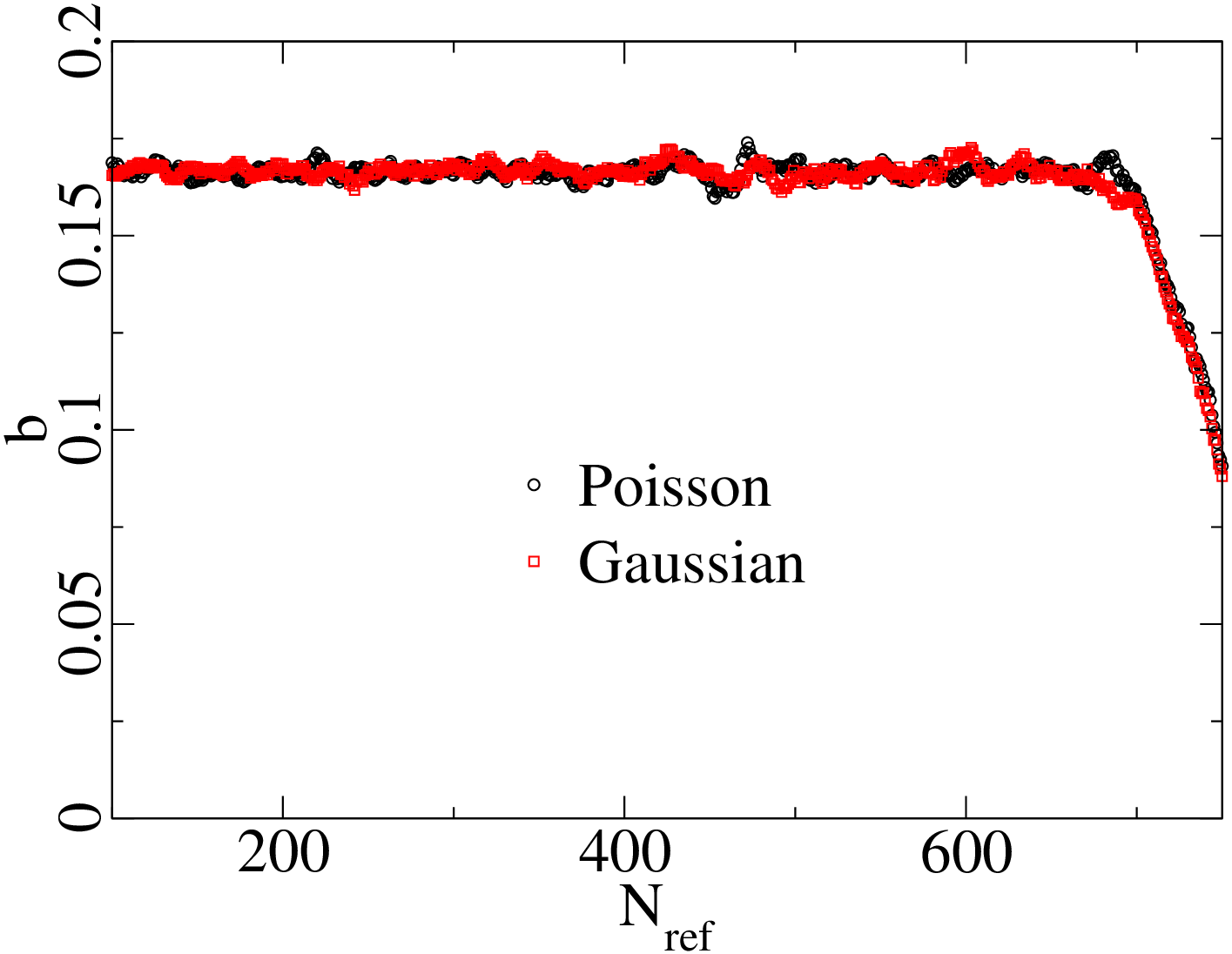}\end{center}
\caption{The correlation coefficient $b$ resulting from a direct Monte Carlo-Glauber
evaluation of the probability distribution \nr{eq:deftorrieri}. Also shown
is the result of the approximation where the Poisson distributions are replaced
by Gaussians, but the distribution of $\npart$ values is the same MCG one,
i.e. \eq\nr{eq:torrierigauss}.
 }\label{fig:torrieri}
\end{figure}

The remaining step is now to integrate \eq\nr{eq:gausstorrieri} over $\npart$. Note that 
this integration is now done for fixed $\nf, \nb, \nref$\footnote{Note that since at no 
point do the experimentalists measure directly the impact parameter or $\npart$, the correct
procedure to get a distribution that corresponds to the measurement 
is to first integrate $P(\npart,\nf, \nb, \nref)$ over $\npart$ to get the 
probability distribution $P(\nf, \nb, \nref)$. From this one can then restrict to
 the conditional distribution 
$P(\nf, \nb | \nref \textrm{fixed} ) = P(\nf, \nb, \nref)/P(\nref)$ 
and compute  $D_{ff}$ and $D_{fb}$. This is \emph{not} 
the same thing as computing $D_{ff}$ and $D_{fb}$
for a fixed impact parameter or $\npart$ and then averaging the results over $\npart$; 
which would imply that $\npart$ is actually measured event-by event.}.
 If, as is typically the case 
for a realistic centrality bin, $\sigma^2_\textrm{p}$ is relatively large
and $\delta_\rmR \gg \delta$, this integral is dominated by the reference multiplicity term
$\left( \nref - \delta_\rmR \tilde{n} \npart\right)^2$. This means that selecting events
with a fixed $\nref$ selects events with $\npart$ close to $\nref/(\delta_\rmR \tilde{n})$,
not the typical value $\langle \npart \rangle$ of the original set of events.
This causes the probability distribution of $\nf$ and $\nb$ to be peaked around
$(\delta/\delta_\rmR)\nref$ and thus induces a correlation between $\nf, \nb, \nref$.
An easy way to 
perform the $\npart$-integration is to realize that the result is a Gaussian
in $\nf, \nb, \nref$. To obtain the coefficients of the 
Gaussian in $\nf, \nb, \nref$ it is then sufficient to evaluate the expectation values and the
covariance matrix, which can also be done by integrating first over the multiplicities and
only then over $\npart$. In any case, from \eq\nr{eq:gausstorrieri} one obtains
\begin{eqnarray}
\langle \nf \rangle &=& \langle \nb \rangle = \delta \tilde{n} \langle \npart \rangle
\\
\langle \nref \rangle &=& \delta_\rmR \tilde{n} \langle \npart \rangle
\\
\langle \nf^2 \rangle - \langle \nf \rangle^2 &=&
\delta \tilde{n} \langle \npart \rangle + 
\left( \delta \tilde{n}\right)^2 \sigma^2_\textrm{p} 
= \langle \nf \rangle + \frac{\sigma^2_\textrm{p}}{\langle \npart \rangle^2} \langle \nf \rangle^2
\\
\langle \nf\nb \rangle - \langle \nf \rangle\langle \nb \rangle &=&
\left( \delta \tilde{n}\right)^2 \sigma^2_\textrm{p} 
= \frac{\sigma^2_\textrm{p}}{\langle \npart \rangle^2} \langle \nf \rangle\langle \nb \rangle
\end{eqnarray}
and correspondingly for the other combinations of $\rmF,\rmB,\rmR$. 
We have how rederived the formulas used in
\eqs\nr{eq:torrierivar} and~\nr{eq:torriericov} to evaluate the result of the 
model of \cite{Konchakovski:2008cf} in the Gaussian approximation previously.

In order to evaluate the importance of the different approximations we have also 
used a simple Monte Carlo Glauber implementation to evaluate the probability
distribution \nr{eq:deftorrieri}. The details of our simple Monte Carlo Glauber model 
are described in Appendix~\ref{sec:mcg} and the result of the computation 
is shown in \fig\ref{fig:torrieri}. It turns out that the result is very close
to $b= \delta/(\delta + \delta_\rmR) = 1/6$. This corresponds to the limit
$\omega_\textrm{p} \to \infty$ in the results \eqs\nr{eq:torrierifb} and~\nr{eq:torrierifr} 
of the toy model. This is easy to understand 
a posteriori. In the Gaussian approximation we assumed that one is only looking
at events in a centrality bin that is defined by a relatively compact 
distribution in $\npart$. This is in fact not the case; the centrality bin is 
defined using the reference multiplicity, and the distribution of
impact parameters $P_\textrm{MCG}(\npart)$ in \eq\nr{eq:deftorrieri} includes the
whole ``min bias'' distribution of impact parameters, and is therefore very wide.
The fact that one must take the limit $\sigma^2_\textrm{p}\to \infty$ means, in
the language of Sec.~\ref{sec:gauss}, that the correlation coefficients 
are all approaching unity. They do so, however, in such a way that 
$b$ approaches $\delta/(\delta+\delta_\rmR)$. This is in some sense the
natural upper limit for $b$, if the correlation function \eq\nr{eq:generalC}
consists of only a Poissonian and long range piece; and the most realistic
way of modifying it is to introduce a short range correlation which increases
the fluctuations in the reference multiplicity. We shall now turn to a more general 
parametrization where this can be seen more explicitly.

\section{The effect of intrinsic correlations}
\label{sec:corr}

\begin{figure}
\begin{center}
\includegraphics[width=0.6\textwidth]{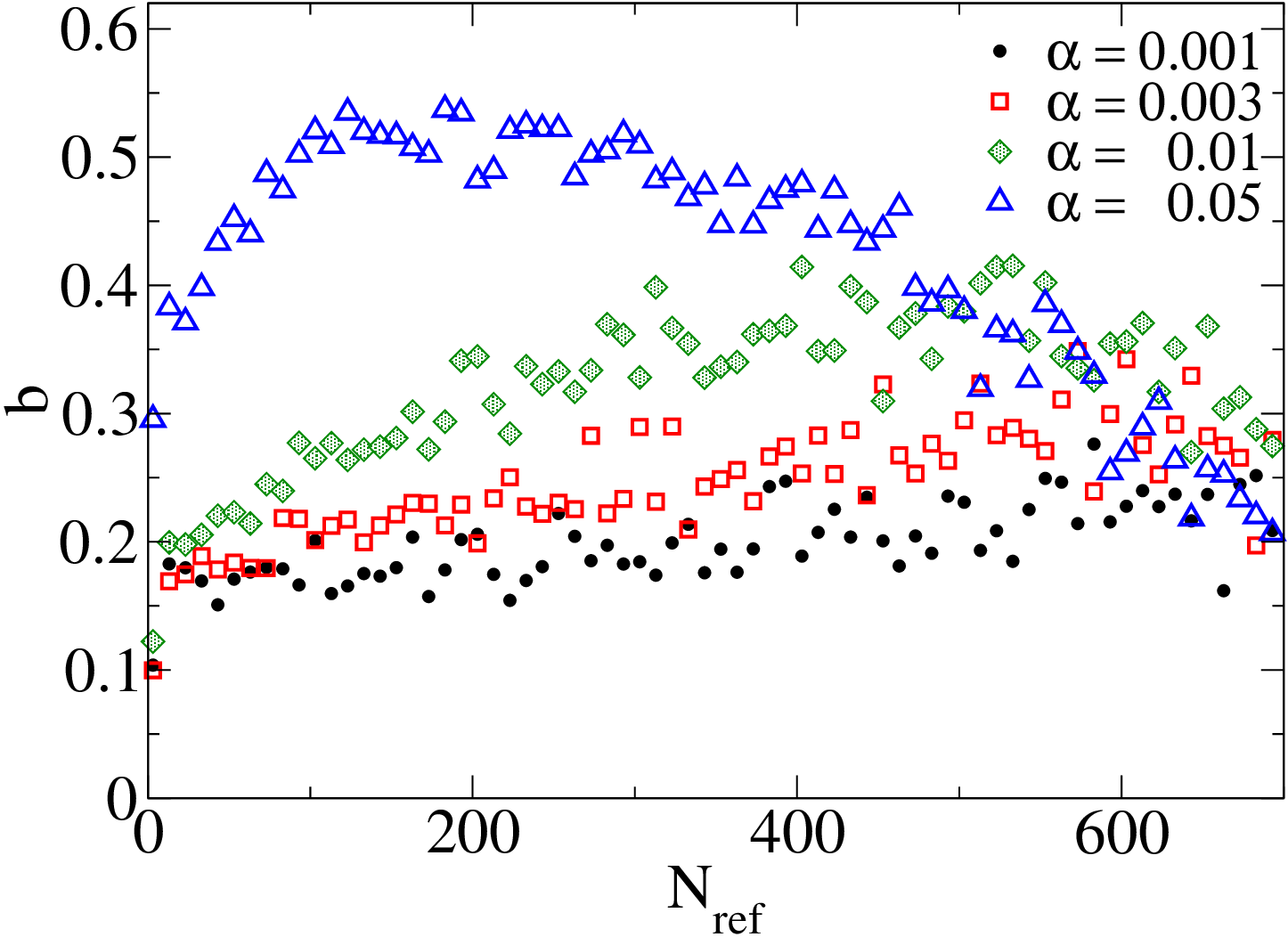}
\end{center}
\caption{
The constrained correlation coefficient $b$ 
as a function of $\nref$ for $K=0.001$ and different values of 
$\alpha$.
 }\label{fig:bK0.001}
\end{figure}

Let us then construct a simple parametrization of long range correlation effects
in addition to the impact parameter fluctuations. As we saw in the previous section,
assuming a Gaussian distribution for the impact parameter fluctuations (i.e.
neglecting the fact that the centrality selection is done using $\nref$) is not
a very good approximation. We would therefore like to construct a model where
the impact parameter fluctuations are parametrized by an $\npart$ drawn from 
a Monte Carlo Glauber calculation, and particle production for a fixed $\npart$
then includes physical long range rapidity correlations.
We emphasize that we are not assuming that the physical mechanism of 
particle production could be decomposed into independent production from 
participant nucleons. We use the quantity $\npart$ as a convenient proxy for the
impact parameter dependent overlap area of the colliding nuclei and its fluctuations.
The main reason for using $\npart$ instead of the overlap area is its easy 
implementation in our simple Monte Carlo Glauber code.

\begin{figure}
\begin{center}\includegraphics[width=0.45\textwidth]{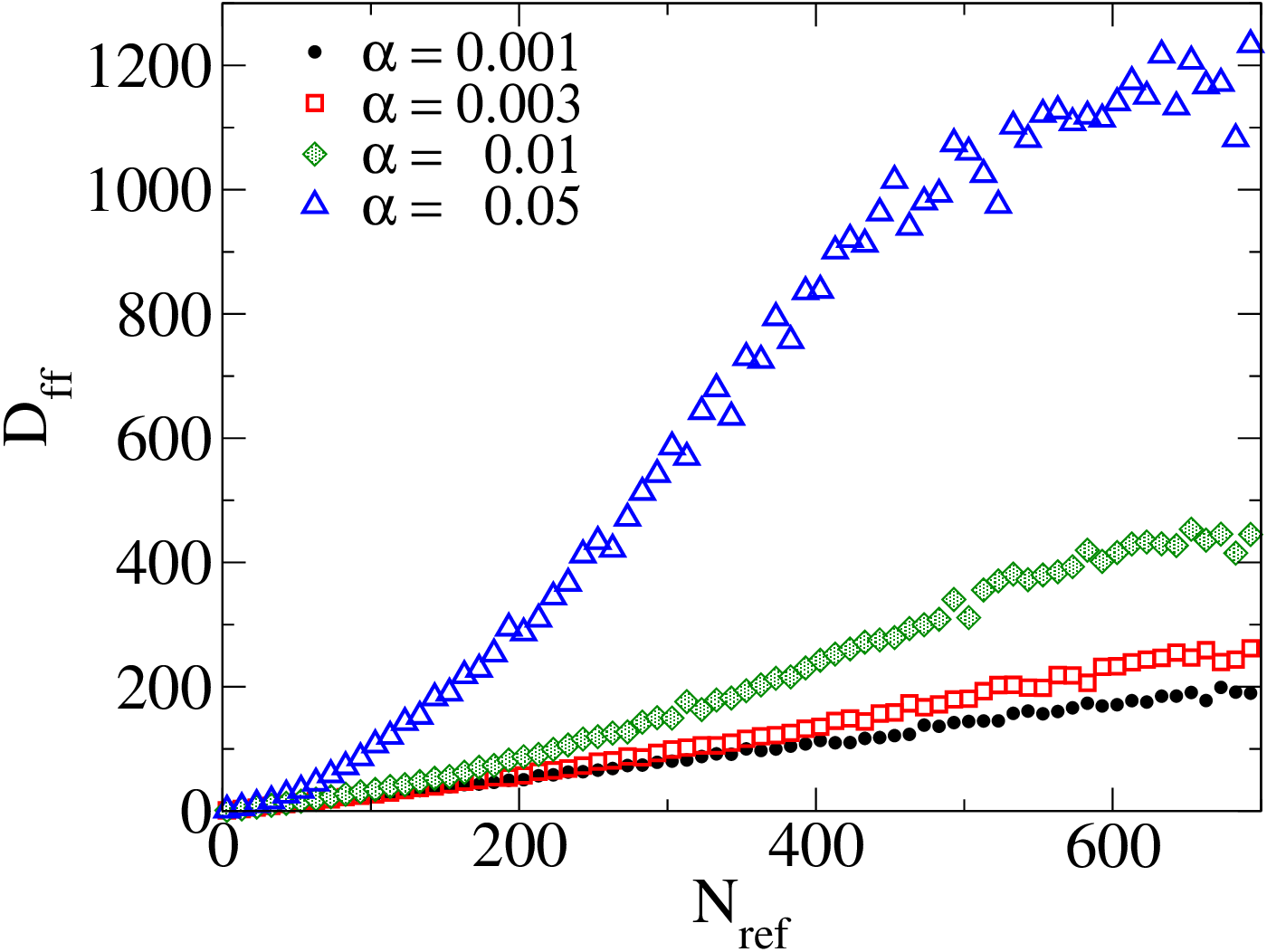}
\hfill
\includegraphics[width=0.45\textwidth]{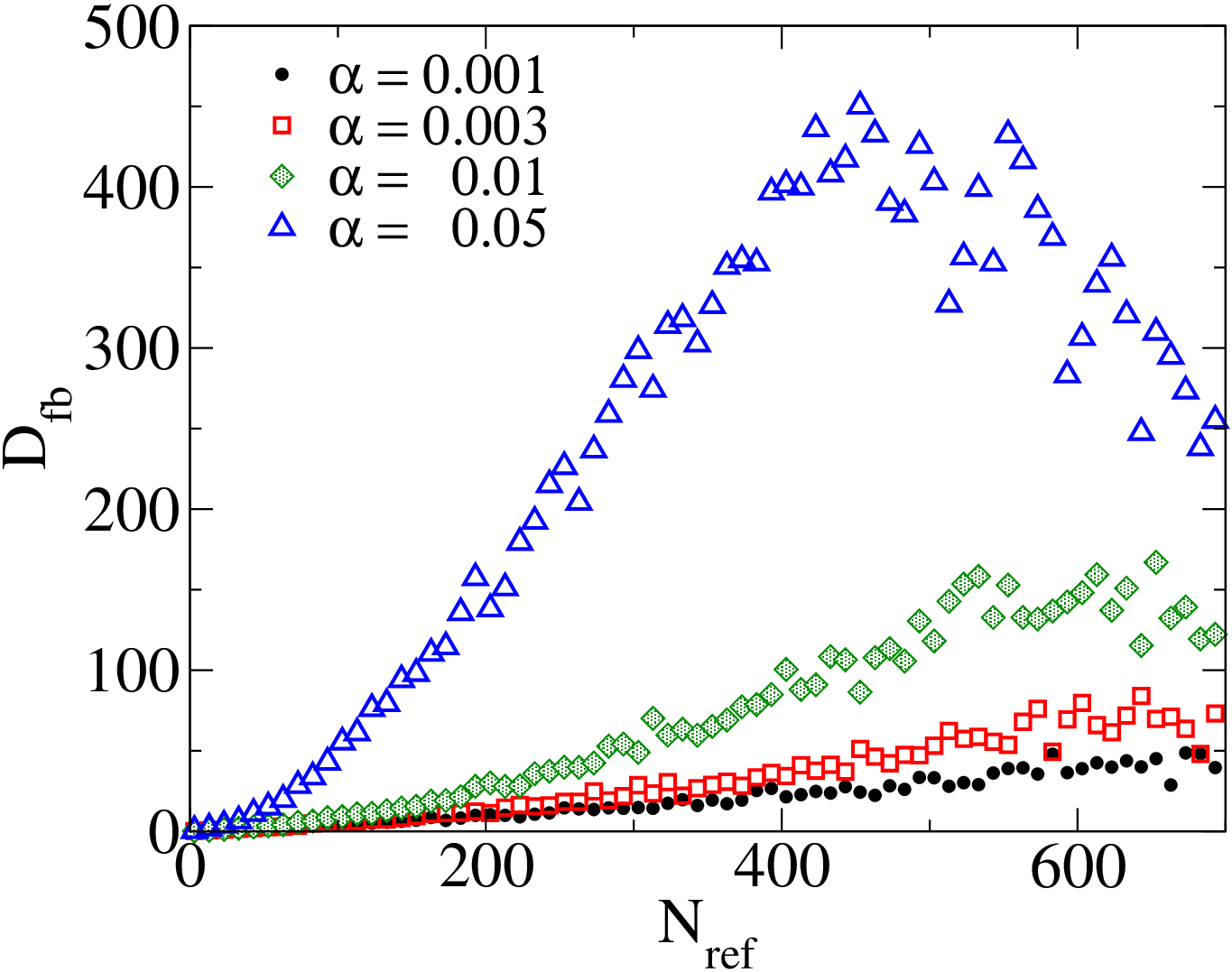}\end{center}
\caption{
The constrained variance $D_{ff}$ and covariance $D_{bf}$ with the same values
for $\alpha$ and $K=0.001$ as in \fig\ref{fig:bK0.001}.
 }\label{fig:DK0.001}
\end{figure}

We will assume that the probability distribution for a fixed $\npart$ 
is Gaussian:
\begin{equation}\label{eq:defparam}
P(\npart,\nf, \nb, \nref) = 
P_\textrm{MCG}(\npart) 
\frac{1}{(2\pi)^{3/2}\det \Sigma(\npart)}
\exp \left[ - \frac{1}{2} \Delta_U \Sigma^{-1}_{UV}(\npart) \Delta_V\right],
\end{equation}
where now $\Delta_U = N_U - \nbar_U(\npart)$ (we denote the expectation
value for a fixed $\npart$ by  $\nbar_U(\npart)$ to separate
it from the measured expectation value $\langle N_U \rangle$ which is averaged over 
some range of impact parameters).
The expectation values 
for a fixed $\npart$ are
\begin{eqnarray}
\nfbar  = \nbbar  \equiv \nbar &=& \delta \,  \tilde{n} \npart
\\
\nrefbar &=& \delta_\rmR \tilde{n} \npart
\end{eqnarray}
and also the correlation matrix for fixed impact parameters depends on $\npart$
through the expectation values $\nbar_U$ as
\begin{equation}\label{eq:defparamsigma}
\Sigma(\npart) = 
\frac{1}{\det \Sigma}
\left( 
\begin{array}{ccc}
\nfbar  + (K +\alpha )\nbar^2 & K \nbar^2  &K \nbar \, \nrefbar  \\[1ex]
 K \nbar^2  &\nbar  + (K+\alpha) \nbar^2 &  K \nbar \, \nrefbar  \\[1ex]
K \nbar \, \nrefbar  & K \nbar \, \nrefbar  & \nrefbar  + (K+\alpha) \nrefbar^2 \\
\end{array}
\right).
\end{equation}
Here we have introduced two additional parameters: $\alpha$ describes the increased 
fluctuations of the multiplicity due to a short range rapidity correlation, and
$K$ represents a dynamical long range fluctuation that increases both the 
local fluctuations $\Sigma_{\rmF\rmF}, \dots $ and generates a correlation 
between the different rapidity windows $\Sigma_{\rmF\rmR}, \dots $. 
In terms of the rapidity correlation function 
these correspond to a parametrization 
\begin{equation}\label{eq:defCparam}
\left. C(\eta,\eta')\right|_{\npart \textrm{fixed}} \equiv 
 \delta(\eta-\eta') \left\langle  \frac{\ud N}{\ud \eta} \right\rangle 
 + \bigg[ K + \alpha \theta(Y-|\eta-\eta'|) \bigg]
 \left\langle \frac{\ud N}{\ud \eta} \right\rangle 
 \left\langle \frac{\ud N}{\ud \eta'} \right\rangle,
\end{equation}
where $Y$ is a characteristic scale of short range rapidity correlations that we assume
is larger than the size of the rapidity windows but less than the separation between 
different rapidity windows\footnote{
We are neglecting the fact that in the actual STAR results the reference window
is often wider than its distance from the F,B windows. Taking this into account 
folly would unneccessarily copmplicate the parametrization \nr{eq:defparamsigma}
without changing our results much.
}. Our parametrization reduces to the model
of Ref.~\cite{Konchakovski:2008cf} in the limit $\alpha = K = 0$. The natural scales
at which the values of $\alpha$ and $K$ can vary can be estimated from the
relation to the parameter $k$ of the negative binomial distribution of multiplicities.
It can be seen from \eq\nr{eq:negbink} that for relatively small 
rapidity intervals  $1/k = \alpha + K$ (recall that this is true for a fixed impact
parameter, or fixed $\npart$ in our parametrization).
For central collisions the PHENIX experiment~\cite{Adare:2008ns} 
$k\sim 690$ for central collisions, so typically we would expect $\alpha, K\sim 0.001$.

\begin{figure}
\begin{center}
\includegraphics[width=0.6\textwidth]{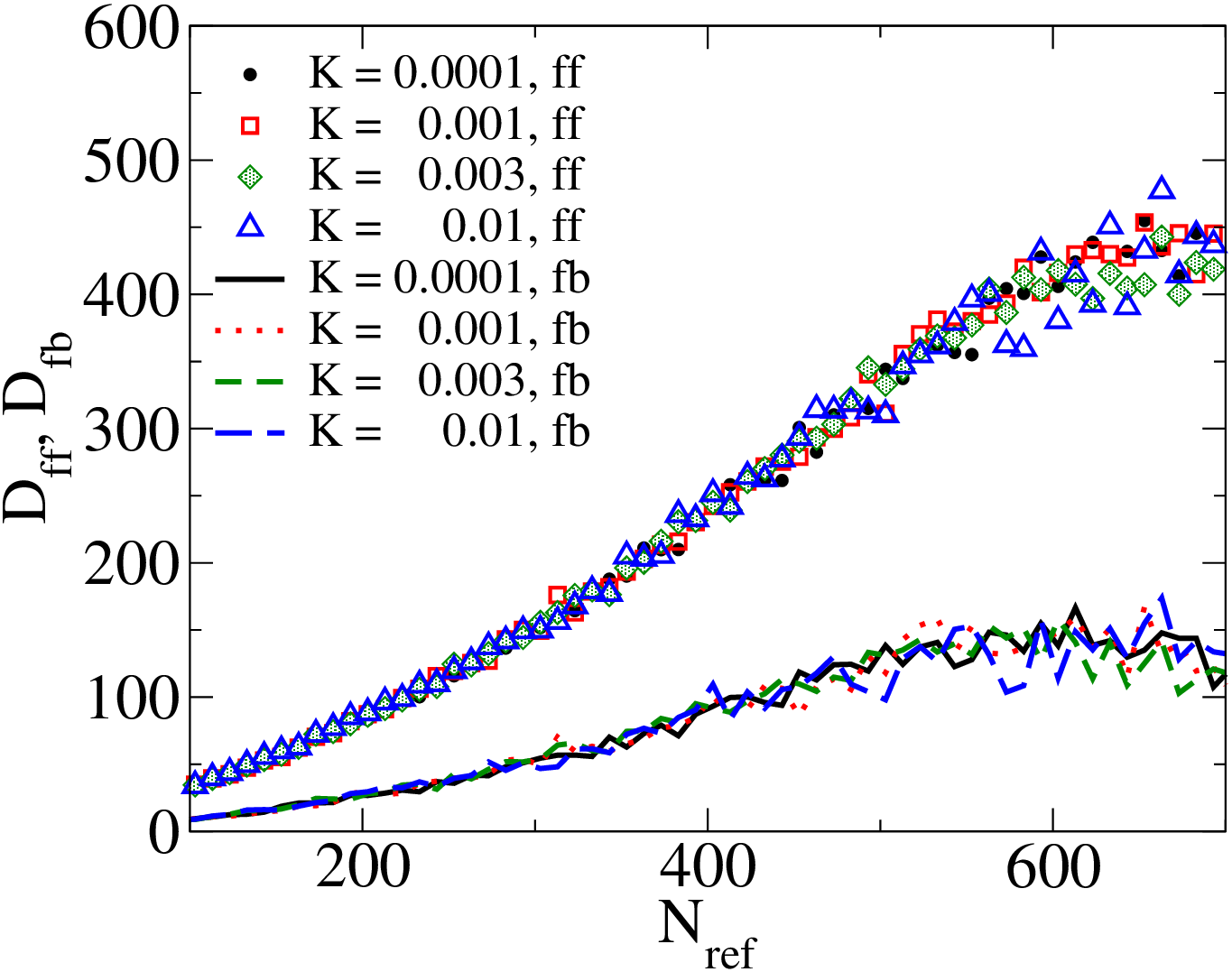}\end{center}
\caption{
The constrained variance $D_{ff}$ and covariance $D_{bf}$ for
$\alpha=0.01$ and different values of $K$.
 }\label{fig:Dalfa0.01}
\end{figure}

\begin{figure}
\begin{center}
\includegraphics[width=0.6\textwidth]{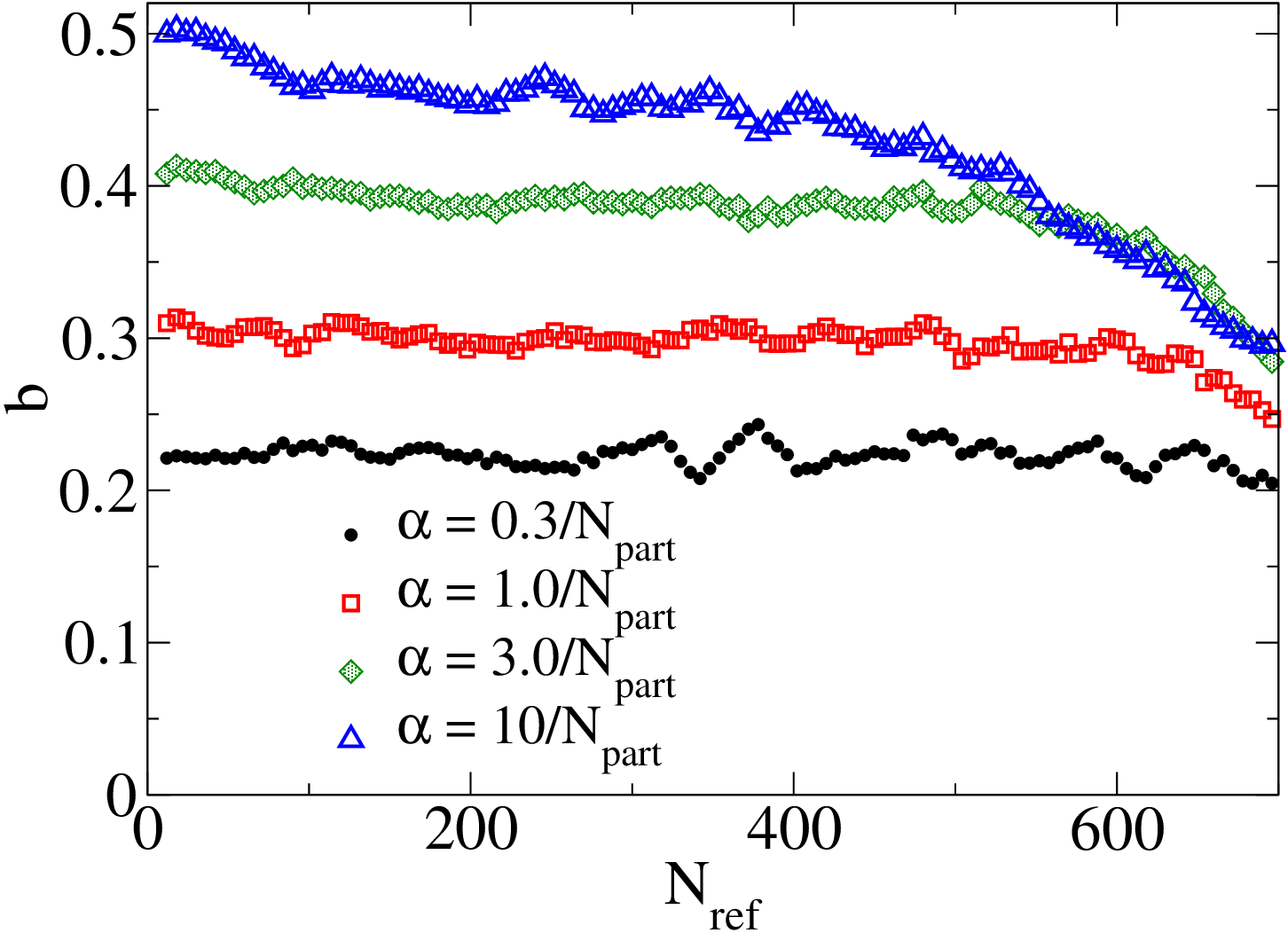}
\end{center}
\caption{
The Monte Carlo result for $b$ with $\alpha$ and $K$ depending on centrality as 
$K=1/\npart$ and $\alpha \sim 1/\npart$ with a coefficient shown in the figure.
}\label{fig:varyak}
\end{figure}

We can then evaluate the correlation using a simple Monte Carlo Glauber implementation 
described in Appendix~\ref{sec:mcg}. Our results are summarized in 
\figs~\ref{fig:bK0.001}, \ref{fig:DK0.001}, \ref{fig:Dalfa0.01}. Figure~\ref{fig:bK0.001}
shows the dependence of $b$ on $\alpha$; the strength of the short range rapidity 
correlation. We see that increasing $\alpha$ increases the correlation $b$. 
This is to be understood in the following way: the main effect of increased short
range correlations $\alpha$ is to increase the fluctuations (uncorrelated
with $\nf,\nb$) in $\nref$. These increased
fluctuations then decrease the correlation between the reference and the F,B 
multiplicities. As can be seen from \eq\nr{eq:b}, this has the effect of increasing 
$b$. For 
a very large $\alpha \sim 0.05$ one can even reach $b\sim 0.5$, but as can be seen 
from \fig\ref{fig:DK0.001} the corresponding values of $D_{ff},D_{fb}$ become 
much larger than the experimental values. Increasing $\alpha$ increases the ratio
$b=D_{fb}/D_{ff}$ closer to the experimental result, but it also increases
the fluctuations in $\nf$ and $\nb$ separately beyond what is observed experimentally.

Figure~\ref{fig:Dalfa0.01} shows the dependence of the correlation on the parameter $K$
describing the dynamical long distance rapidity correlations. It is seen that both
$D_{ff}$ and $D_{fb}$ and therefore also $b$ are, at the statistical accuracy of
our Monte Carlo calculation, independent of $K$. The most natural explanation for
 this is that, as observed previously, the impact parameter fluctuations already 
generate a correlation that corresponds to the limit 
$\sigma^2_\textrm{p} \to \infty$, or a maximal FB correlation. Adding a small
physical dynamical correlation described by a reasonable $K$ is negligible 
compared to this. This leads to a significant observation concerning
the experimental analysis technique, namely that it would be better if the 
centrality selection could be done, into as narrow bins as possible, using another
observable as independent as possible from $\nref$. An example of this effect is 
the STAR observation (fig. 1(a) of Ref.~\cite{Abelev:2009dq}) that determining
the centrality using the ZDC instead of $\nref$ leads to a smaller measured $b$.

In our simple parametrization we have so far completely neglected the dependence of the
parameters on centrality. The charged multiplicity per participant $\tilde{n}$ 
varies among centralities and, more importantly, the dynamical correlation 
strengths $\alpha,K$ should to a first approximation scale with the inverse transverse
overlap area, or $\sim 1/\npart$. Figure \ref{fig:varyak} shows the result 
for $b$ when this kind of a scaling is taken into account.

The purpose of this paper is to stress the qualitative effect of taking into account
the correlation with the reference multiplicity, not to perform a detailed 
fit to experimental data. A further finetuning of the centrality dependence of
the parameters would not change the features we are addressing here.

\section{Conclusions}

The STAR measurements of long range rapidity correlations point to a very 
intriguing picture of strong correlations from the initial strong color
fields in the initial stages of the collision. In spite of the apparent
simplicity of the experimental observable (counting charged
particles in a relatively large region of the detector), the measurements
turn out to be challenging to interpret.
The experimental analysis is done by treating separately events with different 
reference multiplicities. Thus for a consistent treatment one must consider on equal 
footing also the correlations with the reference rapidity window.
This turns the 
problem from a 2-variable into a 3-variable correlation, which has not always
been fully appreciated in the literature. The values quoted by the 
STAR collaboration are for a forward-backward correlation for 
a fixed reference multiplicity.  Since the result shows that for central 
collisions there is a strong correlation between the forward and backward 
multiplicities it would be unphysical to neglect the correlation
with the reference multiplicity.

We have in this paper discussed long range rapidity correlations
in the charged particle multiplicity in terms of only very general 
assumptions on multiplicity correlations. In a Gaussian approximation for the
probability distribution of events in a centrality class, we find an upper
limit $b<1/2$ for the conditional correlation when 
$\delta < \delta_\rmR$ and the F and B windows are far from each other 
in rapidity. We then give up the Gaussian approximation and construct a simple
parametrization of the long and short range correlation, including the effects
of impact parameter fluctuations with a simple Monte Carlo Glauber model.
We show that impact parameter fluctuations alone are not sufficient to explain the 
observed data. Because of the correlation with the reference multiplicity, 
the measured conditional correlation coefficient $b$ turns out not
to be very sensitive to the strength of the dynamical long range correlation.
It does depend strongly on the short range rapidity correlation through its effect
on the fluctuations of the reference multiplicity. Nevertheless we do not find a 
parametrization that would agree with the large $b$ reported by the STAR 
experiment. More experimental data would be welcome to disentangle the interplay 
between impact parameter, short range 
correlations leading to increased fluctuations and genuine long range 
correlations. Having a wide enough rapidity between the measured windows
to actually see the decrease of the long range correlation (as opposed to 
the geometrical ones that are truly infinite range), would be useful but might
not be possible within the STAR TPC.
The dependence of the correlation on the sizes of
the rapidity windows ($\delta$ and $\delta_\rmR$) could be revealing.
Preliminary STAR data indicates that $b$ increases when $\delta$ grows, for a 
fixed $\delta_\rmR$, which is in agreement with our qualitative
expectations based on the  discussion in \se\ref{sec:torrieri}
 Also 
consistency between the STAR~\cite{Abelev:2009dq} and 
PHENIX~\cite{Adler:2007fj,Adare:2008ns} measurements is yet not established, as
they have both a different experimental coverage and analysis method.

 \section*{Acknowledgements}
We are grateful to B.K.~Srivastava for patiently taking the time to answer 
our naive questions and take seriously our suggestions for alteranative ways of 
plotting the data. We gratefully acknowledge conversations and correspondence with 
G.~Torrieri, A.~Bzdak, J.-Y.~Ollitrault, K.~Fukushima, R.~Venugopalan and F.~Gelis.
L.M. thanks the hospitality of CEA-Saclay where this work was initiated as a result of
critical and insightful comments by J.-Y.~Ollitrault.
The research of 
L.M. is supported under DOE Contract No. DE-AC02-98CH10886.  
T.L. is supported by the Academy of Finland, project 126604.

\appendix
\section{MC Glauber implementation} \label{sec:mcg}

We do not use an established Monte Carlo Glauber code, 
but a very simple implementation of our own.
For each configuration we first generate two configurations of nucleons
according to a Woods-Saxon distribution with $\ra= 6.38 \fm$ and surface 
diffuseness $d=0.535\fm$.
We then draw randomly an impact parameter vector $\bt$. If a nucleon is 
at a transverse distance of less than $\sqrt{\sigma_{NN}/\pi}$ with 
$\sigma_{NN} = 41 \mb$ of a nucleon in the other nucleus, it is considered as a 
participant.

Once we have a value $\npart$, one then 
has to generate random variables according to the distribution
\eq\nr{eq:defparamsigma}. For this purpose  one must diagonalize the correlation
matrix \nr{eq:defparamsigma}. This is effectively done in the following way.
One generates 3 independent Gaussian random numbers $\xi,\xi_\pm$ with zero mean
and the variances
\begin{eqnarray}
\langle \xi^2 \rangle & = & 1 \\
\langle \xi_\pm^2 \rangle & = & 1 \pm \frac{K\nbar \, \nrefbar}
{\sqrt{\frac{1}{2}\left(\nbar + \alpha \nbar^2 \right) + K \nbar^2}
\sqrt{\nrefbar + (\alpha + K) \nrefbar^2}}.
\end{eqnarray}
One then solves $\Delta_{\rmF,\rmB,\rmR}$ from 
\begin{eqnarray}
\Delta_\rmF - \Delta_\rmB &=& \sqrt{2}\sqrt{\nbar + \alpha \nbar^2} \: \xi
\\
\Delta_\rmF + \Delta_\rmB &=& \sqrt{\nbar + (\alpha+ 2 K)\nbar^2  }
 \: \Big(\xi_+ + \xi_-\Big)
\\
\Delta_R &=& \frac{1}{\sqrt{2}}\sqrt{\nrefbar + (\alpha+ K)\nrefbar^2  } \:
 \Big(\xi_+ - \xi_-\Big).
\end{eqnarray}
It is straightforward to verify that this procedure gives variables $\nf,\nb,\nref$ with 
the desired correlations,  \eq\nr{eq:defparamsigma}.

\bibliographystyle{h-physrev4mod2}
\bibliography{spires}

\end{document}